# Triple Point Collision and Origin of Unburned Gas Pockets in Irregular Detonations

Yasser Mahmoudi[1*], Kiumars Mazaheri[2]

[1] Department of Engineering, University of Cambridge, CB2 1PZ Cambridge, United Kingdom
Tel.: +44 1223 3 32662; email: sm2027@cam.ac.uk
[2] Department of Mechanical Engineering, Tarbiat Modares University, Tehran 14115-111, Iran

________________________________________________________________

**Abstract**
The turbulent structure of an irregular detonation is studied through very high resolution numerical simulations of 600 points per half reaction length. The aim is to explore the nature of the transverse waves during the collision and reflection processes of the triple point with the channel walls. Consequently the formation and consumption mechanism of unreacted gas pockets is studied. Results show that as the triple point collides with the wall, the transverse shock interacts with the unreacted pocket. After reflection of the triple point off the wall, the transverse wave interacts with the wall. The structure found to be of a double-Mach configuration and does not change before and after reflection. In the second-half of the detonation cell the triple point and the transverse wave collide simultaneously with the wall. The strong transverse wave switches from a primary triple point before collision to a new one after reflection. After some time a weak triple point reflects off the wall and hence the structure exhibits more like a single-Mach configuration. Due to simultaneous interaction of the triple point and the transverse wave with the wall in the second half of the detonation cell, a larger high–pressurized region appears on the wall. During the reflection the reaction zone detaches from the shock front and produces a pocket of unburned gas. Three mechanisms found to be of significance in the re-initiation mechanism of detonation at the end of the detonation cell; i:  energy resealed via consumption of unburned pockets by turbulent mixing ii: compression waves arise due to collision of the triple point on the wall which helps the shock to jump abruptly to an overdriven detonation iii: drastic growth of the Richtmyer-Meshkov instability causing a part of the front to accelerate with respect to the neighboring portions.

________________________________________________________________

## 1. Introduction

Experiments have demonstrated that most self-sustaining detonations are intrinsically unstable with three-dimensional time dependent cellular structure [1]. The lead shock consists of weak incident wave and stronger Mach stem intersect at triple point with the transverse waves, which sweep laterally across the leading shock and collide with each other. The shear layer separates gas streams that have passed through portions of the lead shock with different strengths. The leading wrinkled shock consists of alternate weak incident shocks, stronger Mach stems and transverse waves that are interact at so-called triple points. Two types of detonation structure, weak and strong, corresponding to two distinct types of transverse wave are observed in experiments [1]. In the weak structure, the transverse wave is relatively weak and unreactive emanating from the triple point and extends back into the downstream flow. In strong structure, a portion of the transverse wave close to the triple point can act as a detonation itself that has its own transverse wave. In the previous numerical and experimental investigations, (e.g., [2-9]) both weak and strong types of structure have been observed.

Strehlow and Crooker [7] observed regular structure detonation in hydrogen-oxygen-argon mixture and found that upon collision of a triple point with a wall, the structure was of the weak type. After some time the structure tended to develop a strong type structure. Importantly, the collision and reflection processes of a triple point with a channel wall or another triple point occur in a small region and within very short time. It is therefore extremely hard to capture these experimentally. Hence, numerical simulations are used to properly study the structure configuration in such processes. Furthermore, simulations need very high spatial and temporal resolutions to capture the small structures formed in this problem. So far only few numerical simulations have been performed on the collision and reflection of the triple points. Lefebvre and Oran [3] found that after the collision of a triple point, the structure was a single-Mach configuration with an associated weak transverse wave. However, as the structure evolved it changed to a double-Mach with more complex configuration, where a few kinks appeared along the strong transverse wave. Oran et al. [4] performed two-dimensional computations in a low-pressure mixture of hydrogen and oxygen and suggested the presence of a strong structure of regular detonation. Yet due to low grid resolution the structure configuration around the triple point was not so well identified. Sharpe [6] used high spatial resolution of 64 cells in half reaction length and examined a regular structure of detonation in hydrogen-oxygen-argon mixture. He found that the detonation structure was of double-Mach configuration strong-type with no change in the structure configuration before and after collision of the triple point with a wall. He further commented that a very high spatial resolution was required to clarify the structure correctly. Hu et al. [2] conducted a two-dimensional numerical simulation with resolution of 440 cells per reaction zone length to acquire the detonation structure in regular structure detonation. They concluded that the structure configuration did not change much before and after the collision process, such that the single-Mach configuration appeared after the collision. However, it changed quickly to a double-Mach configuration. They further commented that a sufficiently high-resolution and detailed chemical reaction model are both essential to resolve the structure configuration around the triple point.

In summary, the above numerical simulations determined the structure configuration of laminar detonations characterized by their regular structure. It was found that during the structure evolution process the structure configuration remained unchanged after collision which was double-Mach configuration of strong type. However, both experimental and numerical results indicate the presence of two different types of structure in the mixtures with different activation energies. Smoke-foil technique has displayed irregular transverse wave spacing for high activation energy mixtures and regular cellular structure for low activation energy mixtures [8]. The transverse wave spacing or (i.e. cell regularity) depends on, the mixture composition, initial and boundary conditions [10]. Experiments also revealed that in contrast to the regular structure detonations, there exists an intense chemical activity in the vicinity of the triple point and along the shear layer in irregular structure detonations [11, 12]. Schlieren visualization [13-15] and numerical simulation of the reaction zone in regular structure detonations [16, 17] showed that the shock front ignites almost all the gases that have passed through it. However, in unstable detonations small chemical activity is observed behind the shock front, where pockets of unburned gas escaped from the shock compression and receded further behind the front. Further evidence of two distinct propagation mechanisms and two different structures were obtained from studies on detonation critical limits in porous wall tubes [11, 12]. In irregular structure detonations, re-amplification of the new triple points was observed from instabilities inside the reaction zone. While for regular structure detonations, no re-generation of triple point was observed. Thus, due to such differences it seems that different types of structure configurations arise in different mixtures, when a transverse wave and its associated triple point collide with the channel walls.

Unburned pockets are one of the important phenomena that have been observed both in experiments and in numerical simulations [4, 6, 8, 9, 12-22]. An unburned pocket is defined here as a region of unreacted gaseous mixture that becomes detached from the front and remain behind it. The unburned pockets are speculated to persist to a distance on the order of a cell dimension downstream of the main front. The first evidence of the existence of these pockets was presented by Subbotin [8], who argued that in unstable detonations the transverse waves were unreactive and irregular pockets of unreacted gas were observed behind the front. He described the pockets as triangular with a notch at the vertex. For the stable detonations, however, the transverse waves are reactive and no unburned pockets are formed behind the front. Oran et al. [23] suggested that, the delay in consumption of unreacted gas pockets affects the detonation propagation and may lead to the extinction of detonation. Gamezo et al. [20] by two-dimensional numerical simulation concluded that the burning mechanism of the unreacted pockets varied with the activation energy. They argued that in low activation energy mixtures, the auto-ignition after shock compression consumed the pockets. Nevertheless, in high activation energy mixtures the heat and mass exchange of the neighboring hot gases via diffusion facilitate the ignition of unreacted gas pockets. Sharpe [6] showed that as the triple point collides with the wall, the slip line and the secondary triple point along the transverse wave depart from the front and created an unburned gas pocket. Further, numerical and experimental evidence (e.g. [16, 17, 21, 24, 25]) suggest that in high activation energy mixtures, diffusive phenomena and turbulent mixing, produced by Richtmyer-Meshkov instability (RMI) and Kelvin-Helmholtz instability (KHI) play profound roles in the consumption of the unburned pockets. Nonetheless, the origin of the formation of unreacted pockets, the burning mechanism of these unreacted pockets and, their role in ignition and propagation mechanisms of detonations are not still well resolved.

A very high-resolution numerical simulation is conducted in the present study. The details of detonation structure in high activation energy mixtures are investigated and the following issues are addressed.

i- The structure configuration of an unstable detonation during the collision and reflection processes of the triple point and transverse wave with the channel walls, at the end of the first as well as the second half of a detonation cell.
ii- The mechanism through which the unburned pockets are formed behind the main front.
iii- The influence of hydrodynamic instabilities (i.e. RMI and KHI), upon consumption of these pockets.
iv- The role of transverse wave in consumption of unreacted pocket and consequently the role of these waves in the propagation mechanism of irregular structure detonations.

Thus, the main objectives are:
1. To provide an explanation for the persistence of detonation in the second half of detonation cell, where the weak incident wave is referred to be leading shock.
2. To determine the mechanism responsible for the re-initiation of detonation at the end of the cell cycle, where the shock strength decays dramatically.

## 2. Governing Equations

The current investigation employs two-dimensional reactive Euler equations with a single step Arrhenius kinetics model are. The gases are assumed perfect and the product and reactant gases have the same specific heats. Radiation and other dissipative effects are ignored. The two-dimensional Euler equations are expressed as follows:

$$\frac{\partial U}{\partial t} + \frac{\partial F}{\partial x} + \frac{\partial G}{\partial y} = S \tag{1}$$

where

$$U \equiv \begin{bmatrix} \rho \\ \rho u \\ \rho v \\ \rho E \\ \rho \beta \end{bmatrix} \quad F \equiv \begin{bmatrix} \rho u \\ \rho u^2 + p \\ \rho uv \\ \rho uE + up \\ \rho u\beta \end{bmatrix} \quad G \equiv \begin{bmatrix} \rho v \\ \rho uv \\ \rho v^2 + p \\ \rho vE + vp \\ \rho v\beta \end{bmatrix} \quad S \equiv \begin{bmatrix} 0 \\ 0 \\ 0 \\ 0 \\ \rho W \end{bmatrix} \tag{1-2}$$

Here, $S$ is the source term due to combustion. $u$ and $v$ are velocity component in the $x$- and $y$-directions, respectively. $\rho$, $p$ and $\beta$ are respectively the fluid density, pressure and the reaction progress parameter varying between 1 (for unburned reactant) and 0 (for product). $E$ is the internal energy per unit mas defined as

$$E = \frac{p}{\rho(\gamma-1)} + \frac{(u^2+v^2)}{2} + \beta Q, \tag{2}$$

where, $Q$ is the heat release per unit mass of the reactant, $\gamma$ is the ratio of the specific heats. $W$ is the reaction rate, which follows the Arrhenius law as:

$$W = -k\beta \exp(\frac{-E_a}{RT}). \tag{3}$$

Perfect gas law is expressed as follows,

$$p = \rho RT. \tag{4}$$

The dependent variables are non-dimensionalised with respect to the unburned mixture properties. Density is non-dimensionalised with respect to $\rho_0$, and pressure with $\gamma p_0$. For the velocity, the sound speed of the unburned mixture $c_0$ is used as the reference. The so-called half-reaction length (*hrl*) characteristic is used as the characteristic length scale. This is the length traveled by a fluid particle (in the detonation frame of reference) from the leading shock to the position where $\beta=0.5$ in a ZND structure.

## 3. Numerical method
### 3.1 Numerical solver
A first order splitting method was employed to remove the source term from the system of Eqs. (1), [26-28]. First, the convective terms were addressed by solving the following equation

$$\frac{\partial U}{\partial t} + \frac{\partial F}{\partial x} + \frac{\partial G}{\partial y} = 0. \tag{5}$$

Equation (5) represents the two-dimensional inert Euler equations of gas dynamics in Cartesian coordinates. The system of equations was discretised with the un-split upwind method of Colella [29] and Saltzman [30]:

$$U_{i,j}^{n+1} = U_{i,j}^{n} + \frac{\Delta t}{\Delta x}[F(U_{i-1/2,j}^{n+1/2}) - F(U_{i+1/2,j}^{n+1/2})] + \frac{\Delta t}{\Delta y}[G(U_{i,j-1/2}^{n+1/2}) - G(U_{i,j+1/2}^{n+1/2})], \tag{6}$$

where

$$U_{i,j}^{n} = \int_{\Delta_{i,j}} U(x,y,t^n)dxdy. \tag{7}$$

$F(U_{i+1/2,j}^{n+1/2})$ and $G(U_{i,j+1/2}^{n+1/2})$ are the time averaged approximate fluxes at the cell interfaces. The state variable vectors $U_{i+1/2,j}^{n+1/2}$, and $U_{i,j+1/2}^{n+1/2}$ are determined as the solution of the Riemann problem projected in the *x* and *y* directions with the left and the right states ($U_{i+1/2,j,L}^{n+1/2}, U_{i+1/2,j,R}^{n+1/2}$) and ($U_{i,j+1/2,D}^{n+1/2}, U_{i+1/2,j+1/2,U}^{n+1/2}$). The solution of Eq. (5) was then used as an initial conditions for the systems of ordinary differential equations,

$$\frac{dU}{dt} = S. \tag{8}$$

To generate high-resolution results fine meshes where used locally at the vicinity of the shock front and coarse grids elsewhere. A simple version of the "adaptive mesh refinement" technique of Berger and Colella [31] was utilised. Two sets of uniform grids have been used. The entire domain is covered by coarse grids and fine meshes are superimposed on the coarse grids in the vicinity of the front. This method has been used in 1-D [32] and 2-D [33] simulation of detonation waves. The global procedures of 1-D and 2-D simulations of AMR are similar. The solution at the first cell on the left boundary of fine grids is corrected in order to preserve conservation. The present code has been successfully used for 1-D and 2-D detonation simulations (e.g., [16, 22, 27, 33]).

### 3.2. Boundary and initial conditions
The detonation runs from left to right in the positive *x*-direction in a two-dimensional channel. Since the fluid ahead of the detonation is in its quiescent state, the right-hand boundary condition is irrelevant. The wall boundary condition was imposed on the lower, upper and left sides of the computational domain.

The equations were scaled in such a fashion that the ambient density and pressure are $\rho_0=1$ and $P_0=1/\gamma$, respectively. The half-reaction zone length (*hrl*), is unity, and other parameters took the values $E_a/RT_0=20$, $Q/RT_0=50$ and $\gamma=1.2$ as previously studies in 1D [33] and 2D detonation waves [16, 27]. In this approach *k* should be varied from case to case in order to maintain *hrl* = 1 so *k*= 30.445 for $E_a/RT_0=25$.

Time-dependent two-dimensional calculations were performed for the activation energies 11.93 kcal/mol or $E_a/RT_{ps}$=4.2 (where $T_{ps}$ is the post-shock temperature), characterised as moderately unstable detonations [5, 20]. Other parameters of the reactive system are $\gamma$=1.2 and $Q$=29.79 kcal/mol. For detonations in mixture with $E_a/RT_{ps}$=4.2 the cellular structure has been found to be more irregular [20]. Such detonation structure has been extensively studied numerically (e.g. [2, 5, 6, 16, 18, 20, 21, 34-37]) and experimentally (e.g. [15, 19, 21, 38]). This range of activation energy is the characteristic of mixtures such as hydrogen, acetylene and carbon-monoxide at low pressures ([12, 20, 21, 38, 39]). This range of activation energies which approximately corresponds to the stoichiometric gas mixture $2H_2 + O_2$ at $P_0 = 1$ bar and $T_0 = 293K$ has been studied numerically [20, 40]. It is well established that hydrodynamic instabilities and diffusive processes play profound role in the propagation of detonations in such mixtures [15, 16, 21, 36, 37, 41].

For the initiation of the detonation, a strong blast wave was initiated at $x = 5.0$ which moved to the right and formed a one-dimensional detonation. For a strong blast wave with the shock Mach number $M_s$ at the location $R_s$ with respect to the centre of initiation, the initiation energy is obtained from [33]

$$\frac{E_0}{p_0} \cong R_s \gamma M_s^2 I, \qquad (9)$$

in which '$I$' is a function of only $\gamma$ and $p_0$ is the pressure of the fresh mixture. In order to initiate a detonation a large amount of energy (higher than the critical energy) was deposited instantly in the mixture. In the present work with $\gamma$ =1.2 the value of $I$ was found to be 2.622. In order to have a strong initiation $R_s$ and $M_s$ were chosen equal to 5.0 and 8.925, respectively.

The one-dimensional detonation was then perturbed by adding a disturbance in fresh mixture density between $x$=5.0 and $x$=6.0 [5, 6, 22, 33]. This disturbance has the following form

$$\rho' = \begin{cases} 0 & x < 5.0 \\ 0.25[1+\cos(\pi y/w)]\sin(\pi(6-x)), & 5.0 \leq x \leq 6.0 \\ 0 & x > 6.0 \end{cases} \qquad (10)$$

where, $w$ is the channel width. This perturbation quickly led to the formation of a transverse wave and the two-dimensional detonation forms in the channel. The domain width in the $y$-direction is 6. This permitted to have one detonation modes in the computational domain, which is narrow enough to allow a high-resolution simulation. The computational domain was chosen such that it allowed the detonation to propagate for 400$hrl$, to ensure that the structure is independent of the initial perturbations.

It is well demonstrated that to properly resolve the time dependent structure of the detonations, grid resolution studies should be performed ([5, 6, 16, 42]). In a two-dimensional simulation grid resolutions of less than about 20 cells in the $hrl$ has been found insufficient [6]. It has been also reported that at least 300 cells per half-reaction length ($hrl$) is required to correctly capture the hydrodynamic instabilities (i.e. RMI and KHI) in irregular structures [16]. Whereas, in regular structure detonations 100 cells per $hrl$ constituted good solution in stable detonations [16]. The collision and reflection processes occur at a small region in a very short time step. A very high spatial resolution and small time-step are therefore required to resolve properly all the fine-scales features and structure configuration arising in such process. Hence, in the present numerical simulation resolution of 600 points in the $hrl$ was employed to capture properly the collision and reflection processes of the triple point as well as the formation and consumption mechanism of unburned pocket.

A scalable parallel reactive Euler was developed to carry out the large two-dimensional computations. All calculations were performed on a parallel machine based on the distributed-memory architecture, in which the computational domain is divided into different computational zone and spread among the different nodes. Each node contained processor, which configure as, an Intel® Pentium® 4 with clock speed of 3.00 GHz and up to 1GByte of memory. Communication library of message passing interface (MPI) is chosen for parallelising the code. Typical computation time for resolution of 600 cells per $hrl$ and using double precision accuracy, took about five weeks to run for 400$hrl$.

## 4. Results and discussions
### 4.1. Detonation structure in high activation energy mixture

Figures 1a-d show the contours of pressure, density, temperature and reaction progress variable of the detonation for a mixture with $E_a/RT_0=20$. An extensive description of the structure and the fine-scale features arising in the shock front due the existence of hydrodynamic instabilities, can be found in a previous work [16]. Here, only the main features of the structure are briefly reviewed. These reveal that the detonation front is quite complex, involving interactive shock waves, shear layers and compression waves.

In Fig. 1a the primary triple point (*A*) joins the primary Mach stem (*AM*), the primary incident wave (*AN*) and the primary transverse wave (*Al*). Secondary triple points *B*, *C*, *D*, which have their own transverse waves and shear layers, are clearly visible in Fig. 1a. *Bb*, *Cc* and *Dd* are the secondary transverse waves corresponding to the secondary triple points *B*, *C* and *D*, respectively. The existence of such secondary triple points have been also observed in the previous studies [35, 43]. Figure 1b further shows a jet flow that is developed due to the RMI. *Js* is a shear layer separating the burned gases inside the jet flow from the gases that have passed through the segment *BN* of the incident shock. *BS* is the shear layer of the triple point *B* and separates gasses that pass through the two segments of the incident shock, *AB* and *BN*. Shear layer *DS* separates the gases that have passed through the two sections, *MD* and *DC* of the Mach stem. *VS* is the shear layer associated with the large vortex close to the upper wall that departs the burned gases inside the large vortex with the gases that have passed through the segment *DM* of the Mach stem. *PS* is the primary shear layer, which separates the gases that have passed through *AM* and *AN*. *Al* is the main transverse wave and the kinks *e*, *j* and *g* divide the primary transverse waves into different segments. The straight lines *Ae* and *eg* are the weak non-reactive section of the transverse waves, whose strengths are about 0.48 and 0.72, respectively (Strength is defined as $S=P_2/P_1$ where $P_1$ and $P_2$ are the pressure across the transverse wave, [10]). Line *gj* is the reactive segment of the transverse wave, which its strength is about 1.54 and the interaction of this wave with the primary shear layer results in an explosion at point *j* as seen in Fig. 1a. *jl* is the rare segment of the transverse wave whose strength is about 0.47. Therefore, the detonation structure exhibits a double-Mach configuration of strong type, where a secondary triple point (*g*) exists along the primary transverse wave downstream of the main triple point A.

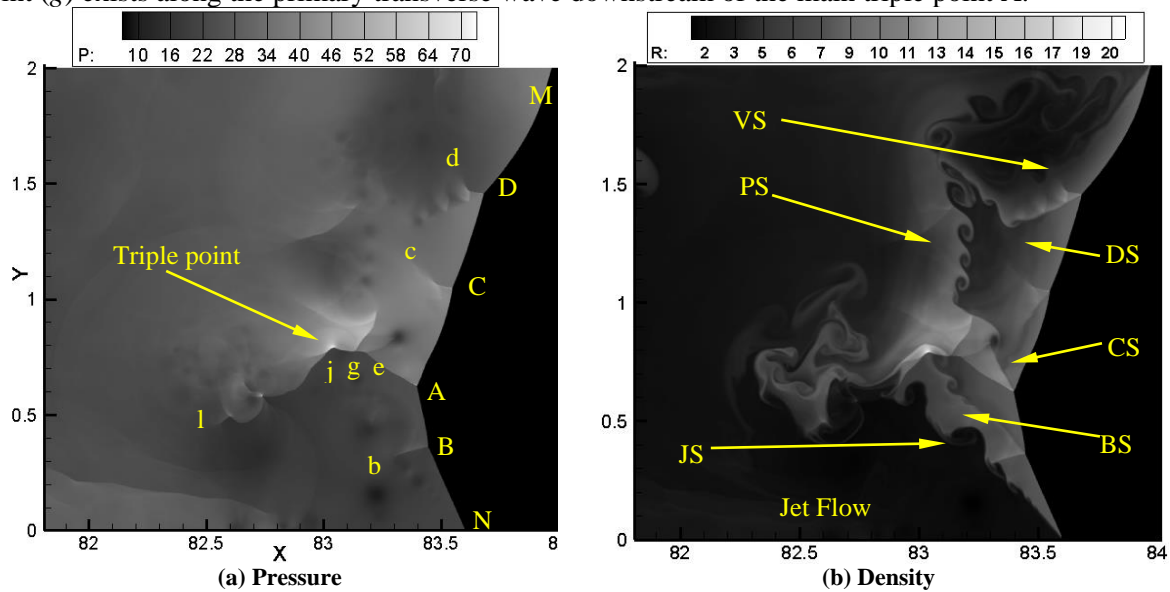

(a) Pressure  (b) Density

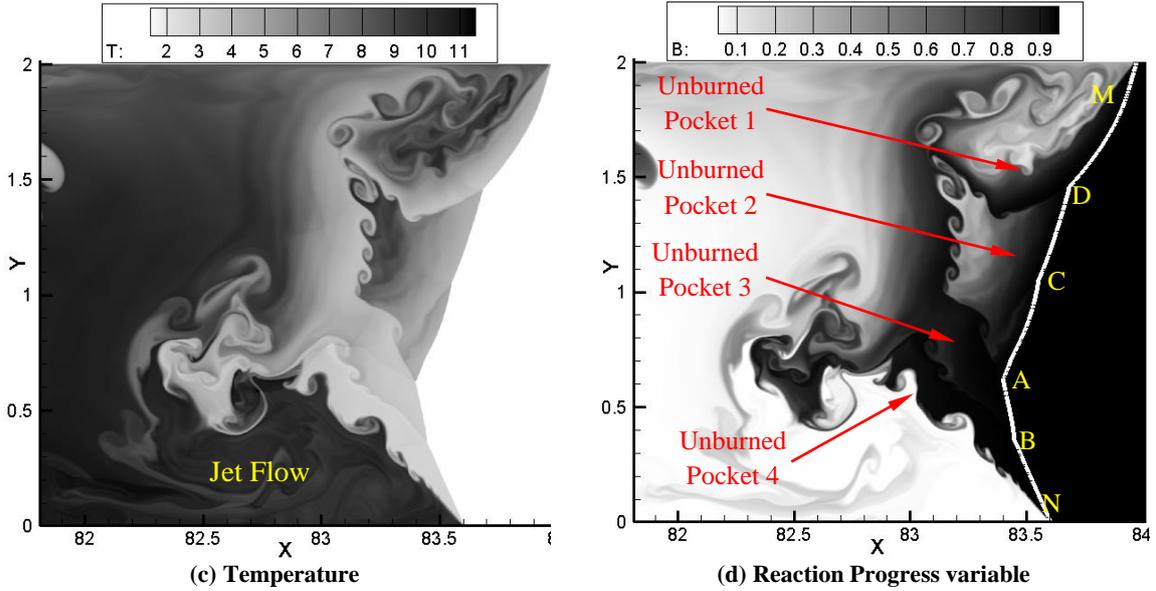

**Fig. 1.** Detailed structure of detonation for mixture with $E_a$=20, $Q$=50, $\gamma$=1.2 and N= 600 cell/hrl. Solid line indicates the shock position.

The reaction progress variable behind the segment *MD* of Mach stem is about $\beta$= 0.91, (see Fig. 1d) which is higher than the reaction progress variable behind the section *DC*, ($\beta$= 0.63) of the Mach stem. Further, Fig. 1c indicates that the temperature of the gases that have passed through segment *MD* is lower than the temperature of gases passed through *DC*. This leads to the formation of unreacted gas pocket (1) behind the shock *DM*, which is surrounded by shear layers *VS* and *DS*. In addition, the unburned pocket (2) with a lower value of $\beta$ than pocket (1) is formed behind the stronger segment *AD* of Mach stem. The two unburned pockets (3) and (4) observed behind the shock front, contain the partly burned gasses respectively passed through the two portions *AB* and *BN* of the incident wave. This shows that in unstable detonations with irregular structure, most portions of the reactant materials escape from the shock compression and remain as unburned gas pockets downstream of the main front. We will explain in the subsequent sections that these pockets eventually burn via turbulent mixing at pocket boundaries due to the existence of vortices produced by the hydrodynamic instabilities.

Figure 2 represents the smoke-foil inscription numerically reproduced based on the peak pressure in the flow field after a fairly long period that the detonation runs with the repeated structure. The cells have convex curvature tracks in the first half-cell and concave tracks in the second half-cell. There is noticeable difference between the structure in the first half and the second half of the cell cycle. Tracks of different triple points, primary and secondary triple points are visible in the first half of the detonation cell. In the second half-cell, however, just two tracks left by the primary triple point *A* and the secondary triple point, corresponding to the transverse wave (*g*), are presents. In the second half-cell, the incident wave sweeps across the channel, hence, the secondary triple points reckoned on this weak incident wave. Thus, in the second half-cell the secondary triple points have less strength than the primary triple points. It follows that, in contrast to the pressure of the primary triple point, the pressures around these secondary triple points are not high enough to imprint any tracks in the second half-cell. While, in the first half-cell the secondary triple points are created on the strong Mach stem. Hence, these points have enough strength, as comparable as that of primary triple point A. Therefore, in the first half-cell the secondary triple points find an opportunity to compete with the triple point (A) and left track on the numerical smoked foil. Thus, in comparison to the upstream portion of the cell, more irregular pattern is observed in the downstream portion of the detonation cell. It may be concluded that the presence of the secondary triple points in the structure, result in exhibition of irregular structure of detonation in high activation energy mixtures.

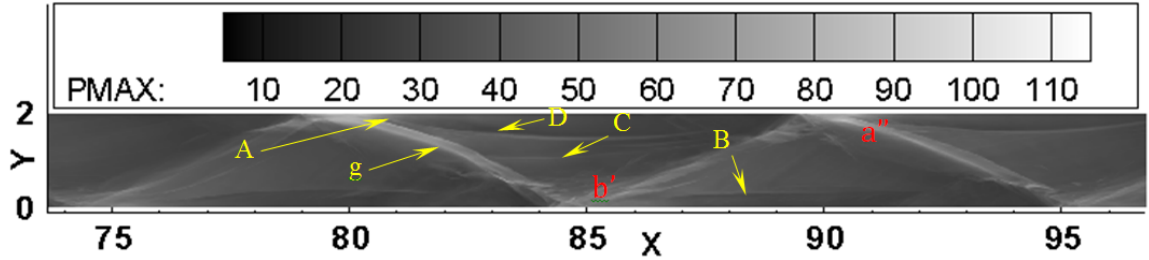

**Fig. 2.** Numerical smoked-foil based on maximum pressure history of detonation in mixture $E_a$=20, $Q$=50, $\gamma$=1.2 and N=600cells/hrl.

### 4.2. Collision with the channel wall at the end of the first half of the detonation cell

This section investigates the structure of unstable detonation during the collision process of the triple point and its associated transverse wave with the wall at the end of the first half of the cell cycle (i.e. collision with upper wall at point a" in Fig. 2). Contours of density and reaction progress variable in Fig. 3 are used to study the interaction of the transverse waves with the shear layers and the formation mechanism of unburned pocket. As Fig. 3a shows, before the collision with the wall the primary triple point A moves upward and is about to collide with the secondary triple point D. Further, the primary transverse wave interacts with the shear layer corresponding to the triple point D. This results in the appearance of a localised high-pressure region ($p\approx50$) behind the front. According to Fig. 3b, the tongue like unreacted pocket with $\beta\approx0.97$ extends back into the hot product behind the main shock. The shear layer corresponding to the triple point D and the shear layer corresponding to the upper jet flow (*jsu*) surround this pocket. Figure 3b further shows that although the transverse wave-pocket boundary interaction produces a high-pressure region such interaction does not alter considerably the pocket burning rate. The jet flow that is produced due to the collision of the triple point with the lower boundary at the end of the last half-cell is also visible in Fig. 3b. This drags the unburned gases inside the primary shear layer into its rolling zone.

Figure 3C shows that prior to the collision, triple point *A* caches up with the secondary triple point *D* and travels as combined triple point *AD* upward to collide with the upper boundary. Further, the shear layers associated to these triple points also merge with each other. Comparison of Fig. 3d with 3b shows that the size of the jet at the lower boundary has increased progressively which indicates the fast growth of RMI. Such increase in the jet size results in dragging of more unburned gas of the primary shear layer into the rolling zone of the jet. Hence, the shear layer has almost vanished in Fig. 3d. Figure 3c further shows that the interaction of the transverse wave with the unreacted pocket scatters the transverse wave into a few weak shocklets. This produces two localized high-pressure region inside the long pocket. The interactions of the reactive section of the transverse wave with *jsu* cause the first pressure rise. The first interaction creates the reflected shock waves *R1* and *R2*. The interaction of shock *R1* with *ps* generates the second high-pressure region inside the unburned gas pocket. Fig. 3d shows that although intense chemical activity is observed at the pocket-transverse wave interaction, the reaction progress variable does not alter noticeably at the interaction points. Further, such interactions do not change even the pocket morphology. Similar behaviours have been also observed numerically and experimentally [17]. Experiments confirmed the existence of intense reactions near the pocket-transverse wave interaction. Further, the interaction of transverse wave with the tongue-like pocket in the numerical simulation did not change the pocket morphology [17].

As shown in Fig. 3e, when the combined triple point *AD* collides with the upper wall a high-pressure region appears at the upper boundary with pressure of $p\approx70$. This is labeled as "3rd high pressure". By this time, the primary transverse wave engages with the unburned gas pocket and does not reach the upper wall. Figure 3f shows that the reaction zone length behind the front at the upper wall is $L_1\approx0.08$ which is two times longer than that at the earlier time, see Fig. 3d ($L_1\approx0.04$). This indicates that when a triple point collides with a wall, the reaction zone decoupled from the lead shock. As the triple point reflects off the wall, the interaction of the reactive section of the transverse wave with the upper wall, causes another high-pressure region at the upper wall and produces a region of $p\approx65$, Fig. 3g. Fig. 3h illustrates that the tongue-shape unburned pocket is isolated from the lead shock and falls further behind the front. This pocket has a triangular shape with a notch at its vertex, which is in excellent agreement with the description of unburned pocket given by Subbotin [8]. Besides, due to detachment of the jet flow from the front reaction zone length behind the front at the upper wall increases, (i.e. $L_1 \approx 0.1$). This further clarifies the decoupling of the reaction zone from the shock front after collision of the triple point with the wall. After reflection, a new primary triple point (*T*), its associated shear layer (*s*) and transverse wave (*Tg*) are formed, Figs. 3i and 3j. The high-pressure region produces a pair of forward and backward facing jets near the upper wall. The interaction of forward jet with the new Mach stem produces a new kink *k*. At the same time, the backward jet moves into the hot and burned gas inside the upper jet flow produced at the last cell cycle. Therefore, the backward jet is consumed which is not apparent in Fig. 3i. The new transverse wave is of strong-type, where the secondary triple point (*g*) and the shock wave *gh* are seen along it, indicating that the structure is like a double-Mach configuration. According to Fig. 3j, the upper jet flow recedes from the front results in the low reaction rate behind the new Mach stem close to the upper wall. Consequently, the reaction zone length increases to $L_1=0.2$. The lower jet flow has grown significantly, which is larger than that at earlier time shown in Fig. 3h. This results in the extraction of unburned gases engulfed in the tongue pocket into its rolling hot zone (see W in Fig. 3j). Figures 3k and 3l show that as the new triple point (*T*) propagates toward the lower wall, the new jet grows progressively. Hence, the reaction zone behind the Mach stem decreases in length. Further, the unburned gases accumulated in the new shear layer (*s*) are dragged into the jet flow. *MD* in Fig. 3k is the weak section of the Mach stem, which is identical to that in Fig1a. Furthermore, the new tongue-shape pockets begins to form behind *MD*. It follows that the genesis of the large tongue-like unreacted gas pockets is behind the Mach stem after collision of triple point at the end of the first half of a cell. During the propagation of detonation, more unreacted gases accumulate inside the pocket that it reaches to its maximum size before the next collision of triple point with the wall at the end of the first half a second cell cycle. Comparison of Figs. 3l and 3j reveals that as the tongue-like pocket recedes more from the front and falls further behind the main shock the jet flow at the down boundary enlarges and the flow luminosity inside it increases. Thus, compared to Fig. 3j, less unreacted gases are dragged into the jet.

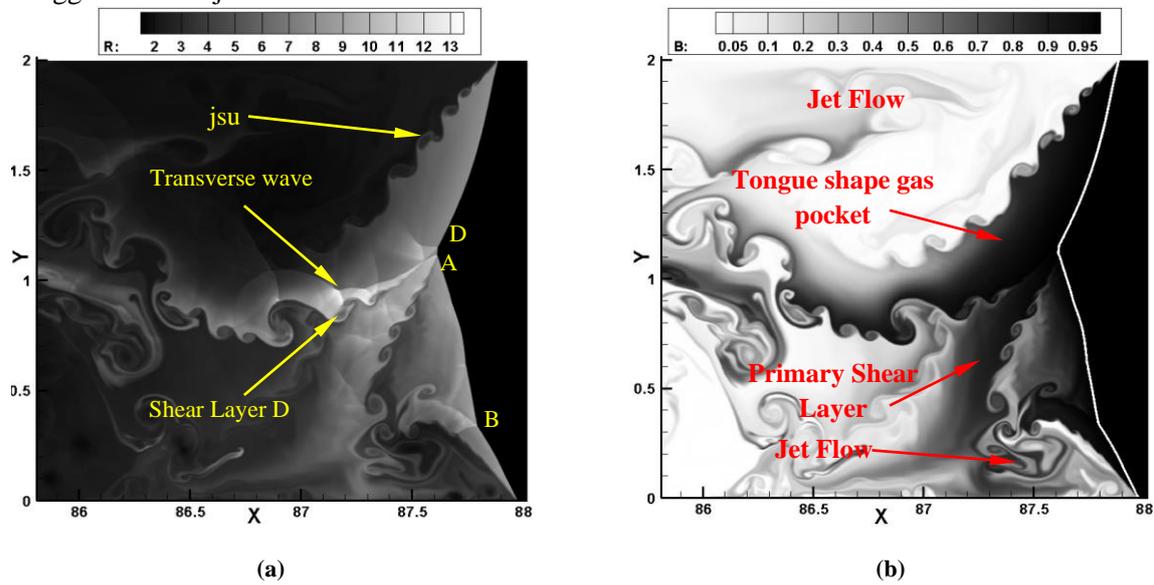

(a)                                                                 (b)

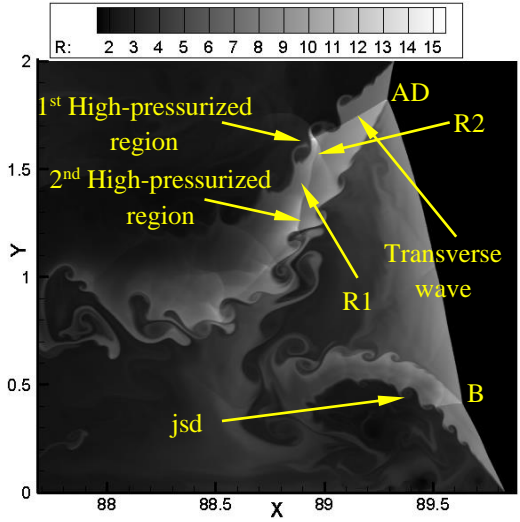
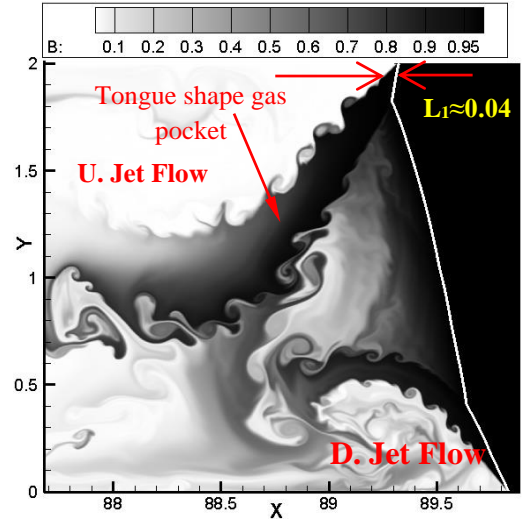

**(c)** **(d)**

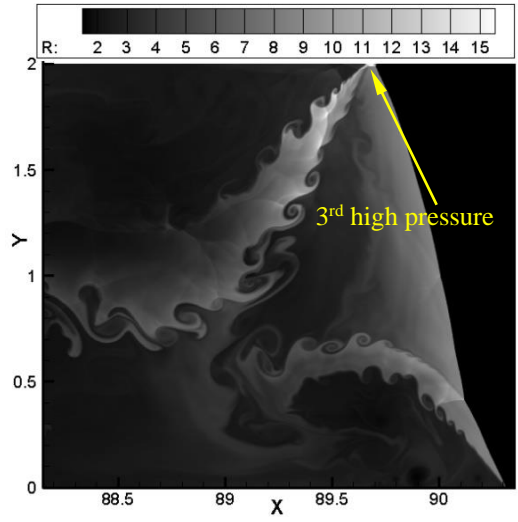
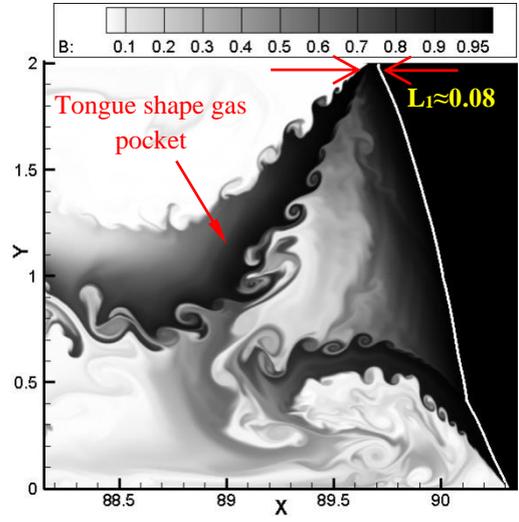

**(e)** **(f)**

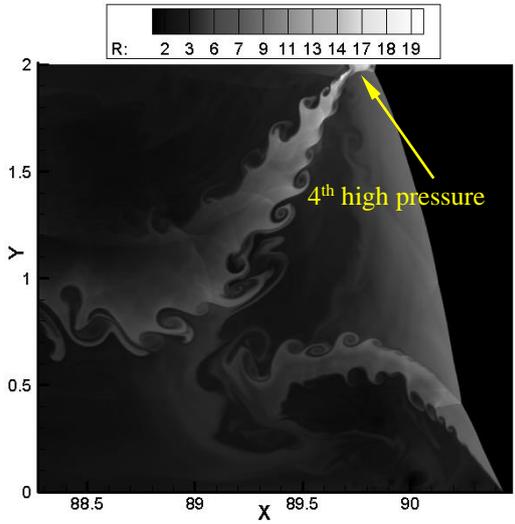
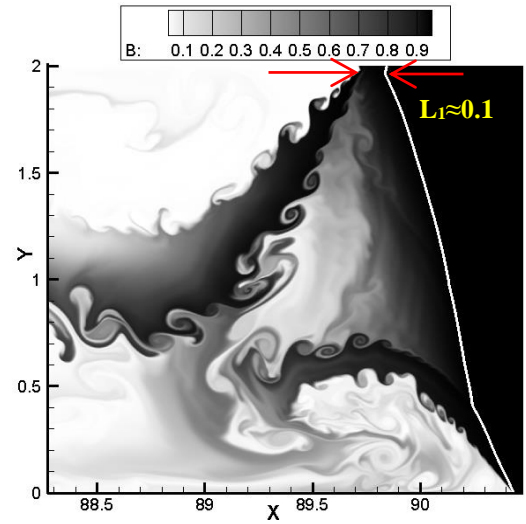

**(g)** **(h)**

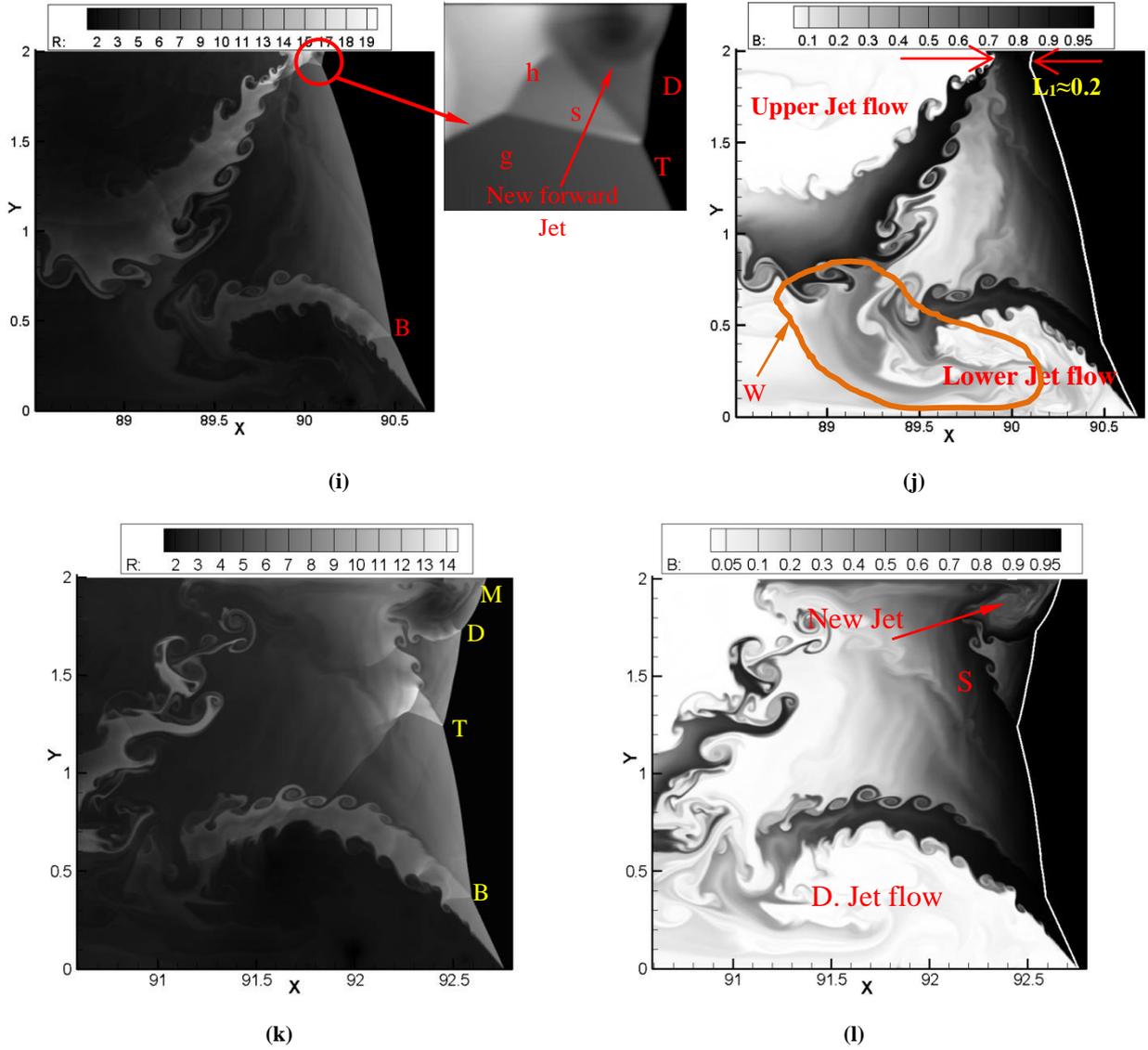

**Fig. 3.** Detonation structure during the collision and reflection processes of the triple point with the wall at the end of the first half of the detonation cell in a mixture with $E_a/RT_0=20$, $Q/RT_0=50$, $\gamma=1.2$ and $N=600\text{cells}/hrl$. Left figures: contour of pressure; Right figures: contour of reaction progress variable. Solid line indicates the shock position.

### 4.3. Collision with the lower wall at the end of second half-cell

In this section the collision and reflection of the transverse wave and its associated triple point with the wall are investigated (i.e. collision with lower wall at point b' in Fig. 2). The contours of pressure and reaction progress variable of the structure. Below the main triple point *T* is the weak incident wave while the upper part is the stronger Mach stem. Figures 4a and 4b represent the detonation structure with one primary and four secondary triple points. In this figure the primary triple point *T* moves downwards and is about to collide with the triple point *B*. The *y* position of the latter point does not change during the propagation of the detonation. Triple point *B* is produced due to the interaction of down jet flow with the front. Since the position of the jet is fixed, the location of the triple point *B* remains unchanged in the structure. The high-pressure region behind the front is produced by the interaction of the primary transverse wave with the shear layer corresponding to the triple point *B* (at point g). This is clearly visible in Fig. 4a. After a while, the triple points *A* and *B* coalesce and move downwards as the triple point *AB* collides with the lower wall, Figs. 4c and 4d. Further, Fig. 4d shows that the shear layers associated with these two triple points merge with each other and create a new shear layer corresponding to the triple point *AB*. Since the transverse location of *B* is constant during the propagation of detonation, the interaction of the triple points *T* and *B* is dissimilar to the collision of two primary triple points. The latter points move towards each other and produce strong explosion. Hence, unlike the collision of two primary triple points such collision results in a lower pressure rise at the interaction point.

Shown in Fig. 4e and 4f are the structure of the detonation when the triple point *TB* collides with the lower wall. The shear layer corresponding to the triple point *TB* merges with the shear layer of the jet flow and create a single "detached shear layer" shown in Fig. 4f. The shear layer is now about to detach from the main front. Due to the interaction of the combined triple point *TB* and its associated transverse wave with the wall, a high-pressure region ($p \approx 72$) appears at the lower boundary. The compression waves resulting from this high-pressure region at the lower boundary are apparent in Fig. 4e. Such high-pressure region was also observed at the upper wall (Fig. 3e) due to the collision of the triple point with the wall at the end of the first half-cell. However, comparison of Fig. 4e and Fig. 3e shows that a larger high-pressure region is produced at the lower wall. This is due to the simultaneous collision of the triple point and the transverse wave with the wall, at the end of the cell cycle. Prior to the collision, the incident shock propagates like an *oblique shock* with a jet flow behind it (see Fig. 4c). However, as Fig. 4e shows the incident shock propagates as a *normal wave* during the collision process. The length of the reaction zone, in Fig. 4d, is about $L_1 \approx 0.11$ while the corresponding length behind the normal incident wave is larger in Fig. 4f, i.e. $L_1 \approx 0.33$. Thus despite the rapid pressure rise at the lower wall producing a region of high temperature and pressure, the reaction rate behind the incident wave decreases. Consequently, the reaction zone length behind this wave increases. This may be contributed to two reasons, as follows:

i- According to Figs. 4e and 4f, it is clear that the high-pressure region emerges inside the hot and burned gases of the jet flow. Hence, it does not affect the burning rate of the unburned gases outside the jet especially behind the incident wave.

ii- As the triple point collides with the wall (Fig. 4f), the jet flow detaches from the front and falls behind it. Hence, the turbulent mixing of the hot and cold gases at the lower wall behind the incident wave decreases. Consequently, the length of the reaction zone behind this wave increases. This emphasises the role of hydrodynamic instabilities, in particular, RMI in the ignition and propagation mechanism of the irregular structure detonations.

Figure 4e shows that the rare section of the primary transverse wave does not interact with the wall. The extended part of the transverse wave is a weak shock wave and therefore its interaction with gas pockets and shear layers scatters it into a system of shock lets. As a result, these shock lets are not strong enough to pass through the unreacted pockets and interact with the wall. By this time the Mach stem becomes much weaker compared to the earlier ones in Fig. 4c. This in turn results in much lower reaction rates behind the wave. This is further evident by comparing the size of the tongue unburned pocket in Fig. 4d which is larger in Fig. 4f.

As illustrated in Fig. 4g the interaction between the compression waves, produced by the high-pressure region, with the normal incident wave forces it to squeeze along the wall. Hence, the incident wave accelerates with respect to the other portions of the shock and propagates as an oblique wave. According to Fig. 4h the "detached shear layer" begins to fall further behind the front creating a new unburned gas pocket. Figure 4i shows that when the incident wave exceeds the other parts of the front a kink (*A*) is formed at the lower part of the shock front close to the lower boundary. This kink is a triple point has its own transverse wave (*AO*) whose strength is about $S=0.91$. It is therefore concluded that before the reflection of a new triple point off the bottom wall a triple point appears in the structure near the bottom boundary. The transverse wave of this point is of strong-type. Thus, this triple point can act as a primary triple point after reflection. Comparison of Figs. 4j, 4h and 4f shows that during the collision process, when the triple point sticks to the wall, the size of the tongue pocket does not change.

After a delay, a new triple point (*L*) appears at the bottom boundary below the triple point (*A*), Fig. 4k. Triple point (*L*) is a secondary triple point and its transverse wave is of weak-type, whose strength is $S=0.31$. The formation mechanism of the triple point (*L*) is related to the following fact: The compression waves produced by the high-pressure region at the lower wall forces hot and rapidly burning gases to squeeze along the wall in a pair of forward and backward jets. As the jets spread, they undergo Richtmyer-Meshkov instability. The backward jet moves into the hot gases inside the detached jet flow (produced at previous cell cycle) and therefore it is consumed quickly. The forward jet, however, moves towards the shock front and interacts with the Mach stem and produces a new kink (*L*), Fig. 4l. An additional shock wave (*gh*) is clearly visible along the transverse wave corresponding to the triple point *A*. Indicating that the structure after the reflection is like a double-Mach configuration of strong-type. After reflection the lower part of the main front, below the main triple point (*A*), becomes a Mach stem while the upper part is an incident wave. Hence, the tongue-like pocket is placed behind the weak incident wave. Thus, more shocked unreacted gases, passed through the incident wave, accumulate in the pocket. Therefore, the size of the pocket increases significantly. This can be deduced through comparing the size of this pocket in Fig. 4j with that in Fig. 4l.

The sequence studied above reveals that in the detonations with irregular structure, at the end of the detonation cell, a double Mach configuration structure interacts with the wall. After reflection a weak triple point appears, and the structure resembles a single Mach configuration. However, prior to the formation of this weak triple point a strong triple point forms near the wall, whose transverse wave is of strong-type. In other words, before the reflection of a new triple point off the wall, the strong transverse wave switches from a main triple point prior to the collision to a new primary triple point after reflection. Hence, the structure has a double-Mach-like configuration of strong type and there is no change before and after collision of the triple point.

Shown in Figs. 4m and 4n are the detonation structures at the moments when the main triple point (*A*) moves further away from the wall. The secondary triple points, *g*, *D* and *L* are clearly visible in Fig. 4m. Further, the new forward jet is now much larger than that in earlier times. Hence, the reaction rate behind the Mach stem increases, consequently the length of the reaction zone behind it decreases. Furthermore, the weak incident shock continuously engulfs gases with very long ignition delay time. This can be seen from the increasing size of the unburned layer in Fig. 4n. The layer separates the burned gases inside the upper jet flow from that processed by the incident shock. This unreacted pocket extends back into the hot and burned gas behind the front. In addition, the flow circulation inside the upper jet flow drags the unburned gases into it and facilitates the ignition of the gas. The existence of such tongue-like unreacted gas pockets has been previously confirmed experimentally and numerically [16, 17].

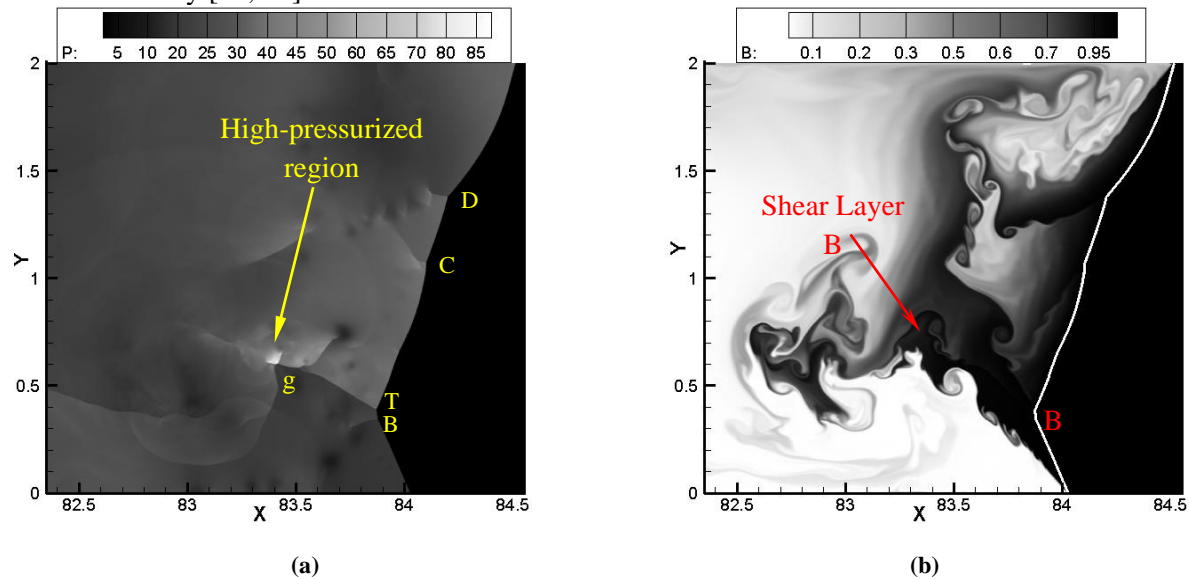

(a)   (b)

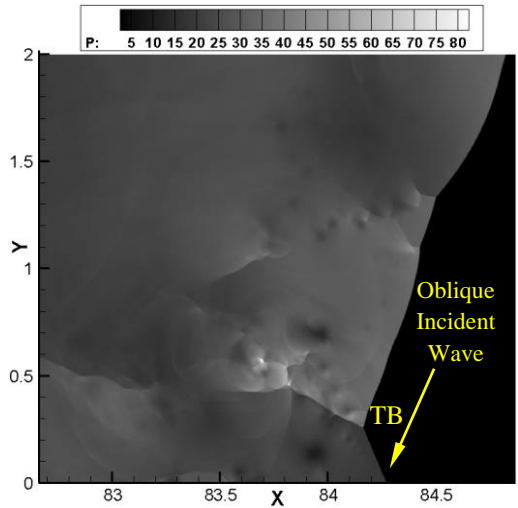
(c)

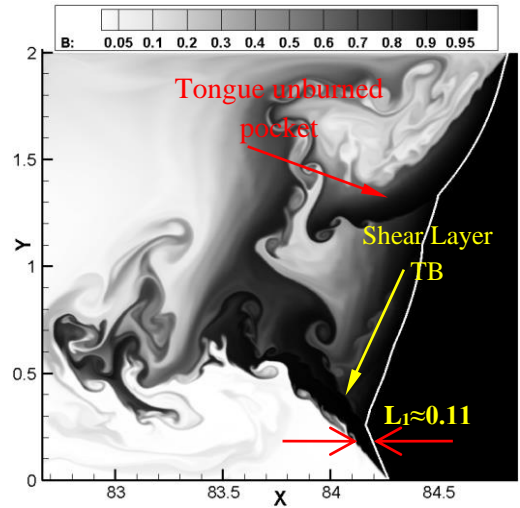
(d)

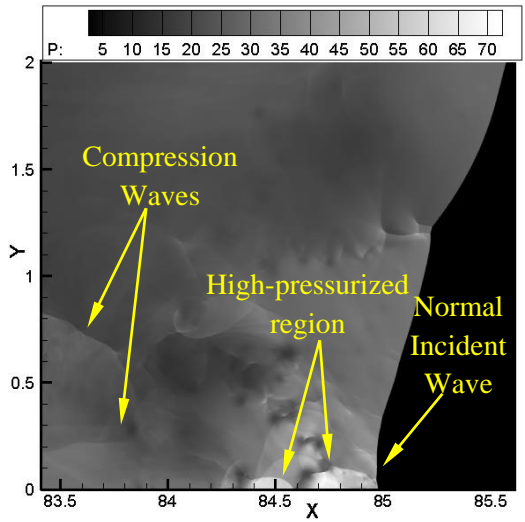
(e)

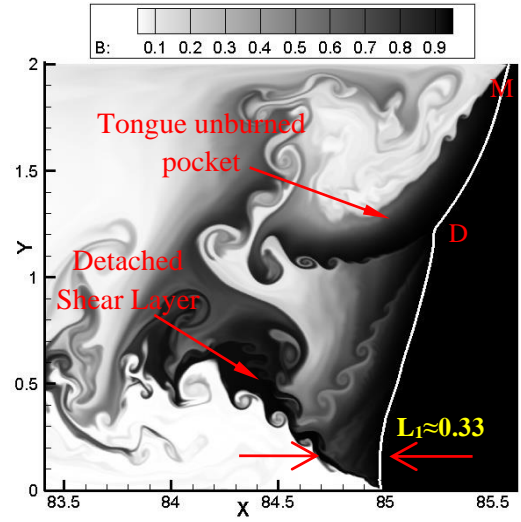
(f)

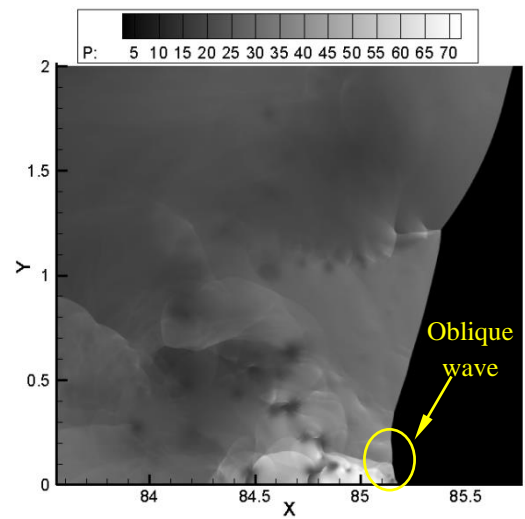
(g)

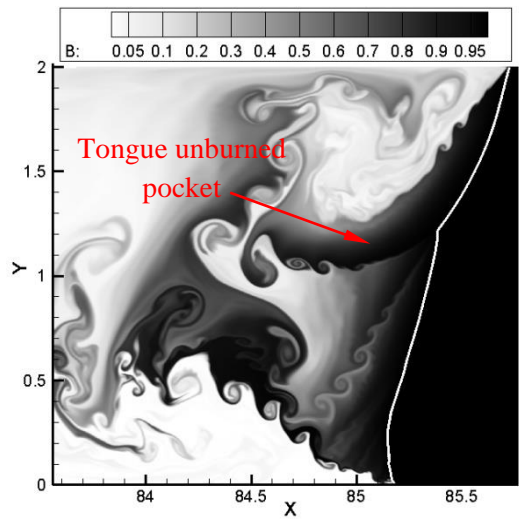
(h)

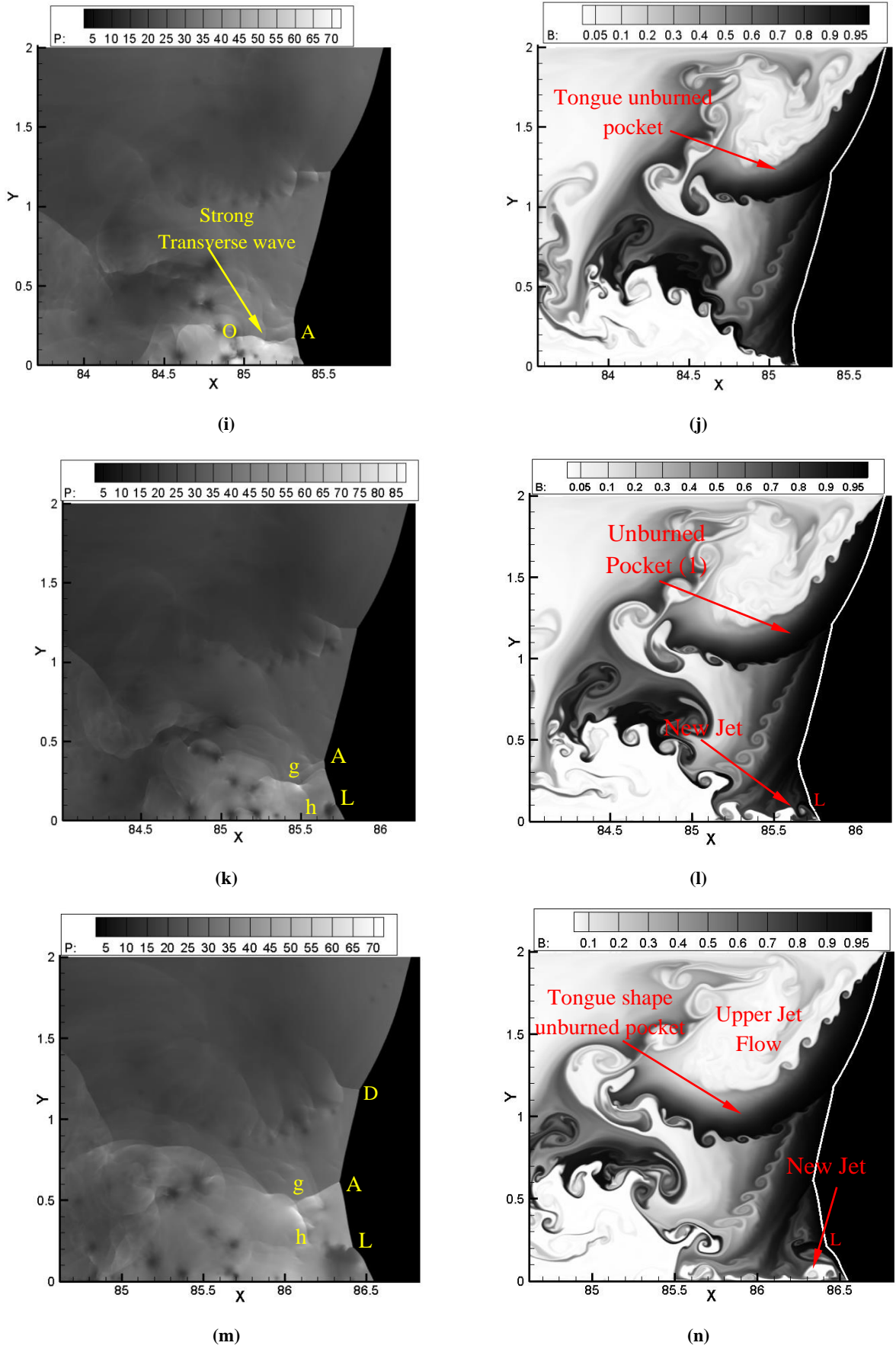

**Fig. 4.** Detonation structure during the collision and reflection processes of the triple point with the wall at the end of the second half of the detonation cell in a mixture with $E_a/RT_0=20$, $Q/RT_0=50$, $\gamma=1.2$ and N=600cells/$hrl$. Left figures:

contour of pressure; Right figures: contour of reaction progress variable. Solid line indicates the shock position.

### 4.4. Tongue-like unreacted pocket and hydrodynamic instabilities

Figure 5 shows a series of snapshots illustrating the detonation structure in the second half-cell. These correspond to the travelling time of the triple point from the upper wall to the down boundary. As illustrated in Fig. 5a, the tongue-shape pocket is isolated from the front after collision of the triple point with the upper boundary at the end of the first half-cell. As the triple point moves downward (Figs. 5c and 5d) the size of the pocket decreases. By the time the triple point reaches the lower wall, Fig. 5e, almost all the gases inside the pocket have been consumed. Hence, the pocket remains for half a cell behind the front. This is in keeping with the existing experimental works 21][ which state that the tongue-like unreacted gas pockets will burn during a half cell cycle. According to Fig. 5 the jet flow near the down boundary, which goes through the Richtmyer-Meshkov instability, drags the capsule of the unreacted gas into its rolling zone and facilitates the burning rate of the pocket. Further, the secondary instability, i.e. Kelvin-Helmholtz instability, also develops along the large vortex boundary and causes vortex roll-up in small-scales. Ultimately, a turbulent mixing zone develops between hot and cold gases. Nonetheless, as Fig. 5 shows the vortices produced by RMI are significantly larger than KHI vortices. Furthermore, as shown in Fig. 3i due to strong circulation flow field inside the jet, most portions of the unreacted pocket moves into the jet flow and burn in this region. It may be concluded that the role of RMI in turbulent mixing and consumption of unburned pockets is more important than KHI. The interaction of transverse waves with the pocket boundary causes localised explosions inside the pocket (see Fig. 4c). However, these explosions alter neither the morphology nor the burning rate of this pocket. A consequence of the transverse shock interaction with the pocket boundary is the shock scattering of the transverse shock into a system of shocklets (R1 and R2 in Fig. 4c). It follows that the transverse waves in irregular structure detonations are of strong-type. However, these waves are not strong enough to contribute directly with the ignition and propagation mechanism of gaseous detonations. Instead, they contribute with the formation of large vortices by collision of their associated triple points with the side walls. These have substantial role in the ignition of unstable detonations.

It is well demonstrated that the variation in the range of scales with activation energy appears to be a consequence of the Arrhenius dependence of the reaction rates and the lead shock velocity oscillation. The very long induction time caused by the very low shock velocity ($V=0.8V_{CJ}$) at the end of the cell, may lead to local decoupling of the lead shock and the reaction zone [14, 15]. However, so far, the re-initiation mechanism of detonations at the end of the decaying portion of detonation cell has not been resolved well. These results reveal that the energy released by the consumption of the tongue-shape pocket within the second half of the cell, augments the lead shock front and helps the self-sustained propagation of the detonation. Furthermore, a strong explosion occurs at the lower boundary. This is caused by the simultaneous collision of the triple point and the transverse wave with the wall and produces compression waves, which support the shock front. This can be a reasonable mechanism responsible for the re-initiation of detonations at the end of the cell cycle. It is, therefore, concluded that the hydrodynamic instabilities and the transverse waves are of significance. The former has a direct role while the latter plays an indirect role in the detonation propagation and re-initiation mechanism in unstable detonations.

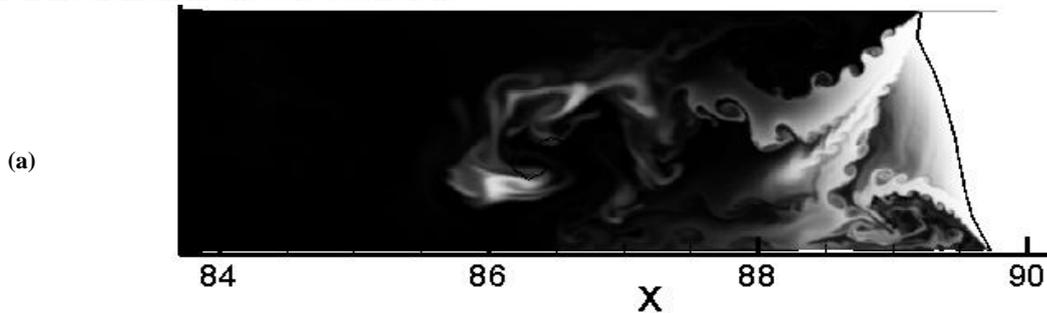

(a)

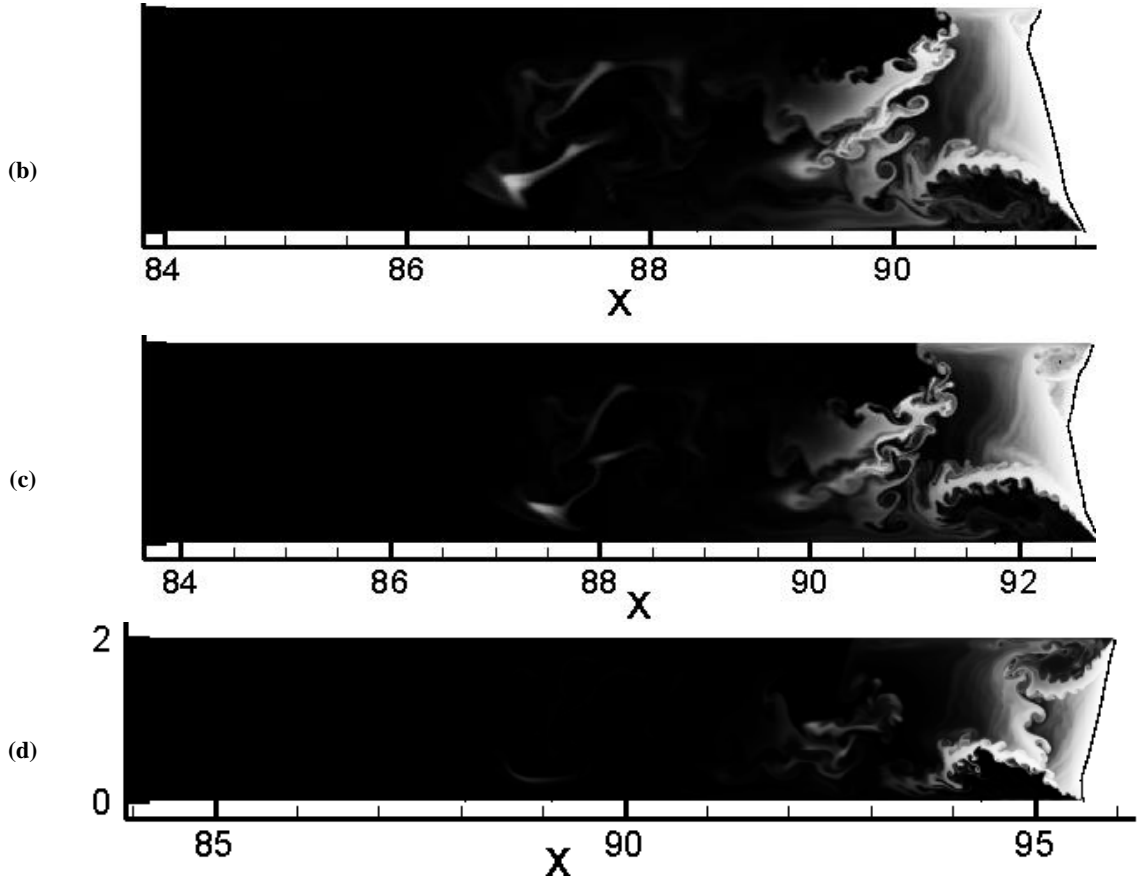

**Fig. 5. Contour of reaction progress variable illustrating the detonation structure in the second half of a detonation cell.**

### 4.5. Diffusion effects

In the present work, we performed Euler simulations. Such simulations are not convergent under low resolutions since the numerical diffusion is proportional to grid spacing. Perhaps the Euler equations are not still an adequate physical model at this level of resolution and the Navier-Stokes equations are necessary to resolve the real diffusion terms. Mazaheri et al. [37] performed Navier-Stokes simulations and compared the obtained structures by those obtained through Euler equations in both regular and irregular structure detonations. This revealed that due to the absence of hydrodynamic instabilities and un-reacted gas pockets in regular structure detonations, the structure captured by solving these equations are very similar in low activation energy mixtures. In the high activation energy mixtures the diffusion suppresses both the small-scale and large-scale vortices produced by KH and RM instabilities. Nonetheless, qualitatively the solution of the Euler and Navier-stokes equations are similar. Furthermore, previous investigation (e.g., [24]), by one-dimensional simulation, showed that the influence of diffusion at the shear layers depends on the activation energy. This is negligible for regular structure detonations. Further, two-dimensional simulation in high activation energy mixtures [44] have shown that physical diffusion is important at high resolution when the numerical diffusion is negligible. Hence, for accurate detonation wave solutions it is necessary to solve full reactive Navier-Stokes equations. However, their results showed that the structures obtained by solving the Euler equations and Navier-Stokes equations were qualitatively similar. Hence, we come to the conclusion that the results described here at resolution of 600 points per hrl do not qualitatively change as one goes to higher resolutions.

### 4.6. Three-dimensional effects

The obtained numerical results on when and how the unreacted pockets burn, agree reasonably well with the experimental observations. There is, however, a concern regarding the consumption time of the unburned pocket in two-dimensional numerical simulations compared to that in experiments. There is further an uncertainty about the similarity of the results in the two- and three-dimensional calculation. This is due to the nature of the turbulence in the flow field behind the main shock. Turbulence is generated in detonations by the baroclinic mechanism which occurs during the interaction of the compression waves with the high-density gradients along the shear layers. This leads to the vorticity generation on a large scale comparable to the scale of the system. Another source of turbulence on the small scales is the KHI. However as indicated by the results, it appears to be less significant than the RMI for the flows under investigation. Radulescu et al. [17, 21] suggested that such mechanism as a source of turbulence behind the front in gaseous detonations. This type of turbulence (shock-flame interaction) has also been observed in DDT phenomena (e.g. [45-52]). Oran and Gamezo [51] showed that the computations in two dimensions were in qualitative agreement with those in three dimensions and experiments. Gamezo et al. [46] pointed out that though 3D simulation the flame becomes more wrinkled, however, the overall flame development is still dominated by RMI. Hence, their two-dimensional and three-dimensional simulations showed very similar results.

Furthermore, the behaviour of RMI have been isolated and studied by many authors, (e.g., [53-56]). The numerical study of RMI, performed [57] have shown that the growth rate of perturbations in the 3D case was higher than that in 2D with the identical initial conditions. Li and Zhang [56] compared the two and three-dimensional temporal growth rate of RMI. In the non-linear regime, the growth rate in three-dimension was about 20% larger and 25% faster than that in two dimensions. However, in the linear regime the growth rate of the instability in the two and three dimensions were the same.

Here the RMI is the dominant mechanism for the appearance of turbulent structure behind the shock front. This is expected to be the reason for the qualitative agreement between the present two-dimensional results with those in three dimensions and experiments. However, to quantify the results and determine the accurate consumption time of the unreacted pockets, three-dimensional computations should be performed. Further, it should be noted that, in detonation waves, the nature of the turbulence differs from the classical Kolmogorov turbulence. Further, compressibility effects are of significance in RM instabilities and chemical reactions provide energy in maintaining the turbulent fluctuations on all scales [11, 21]. Such turbulence occurring in all scales is referred to as non-equilibrium turbulence or non-Kolmogorov turbulence [51]. Thus modelling of turbulence in the unstable detonations through using classical turbulence model, introduces difficulties. This is because turbulence at fine scales play a significant role in the gas ignition mechanism. Hence, a closure model should be proposed to describe all the complexities arising in the non-Kolmogorov turbulence. The properties of this model and how it relates to the classical is still an unresolved problem upon the detonation community.

## 5. Conclusions

A high-resolution Euler numerical simulation was conducted to determine the detonation structure evolution during the collision and reflection of the triple point with the wall. This includes the end of the first half of detonation cell as well as at the end of the decaying portion of the cell cycle. The major contributions of this work are as follows.

i. At the end of the second half of detonation cell, the primary triple point and the transverse waves collide simultaneously with the wall. In the first half-cell, when the triple point reaches the wall, the transverse wave interacts with the tongue-like unreacted pocket. Thus, in comparison to the first half-cell, much larger high-pressure zone is generated at the end of the second half-cell.

ii. Due to the strong explosion at the lower wall (end of the second half-cell), the incident wave accelerates relative to the neighboring parts of the front. This produces a kink in the front close to the lower wall. This kink has its own strong-type transverse wave where the secondary triple point and reflected shock along it are produced. After some time a new triple point reflect off the wall which is of weak-type. Therefore, it is concluded that after collision with the lower wall the transverse wave switches from a strong triple point before collision to a new one after reflection.

iii. Due to the weak explosion at the upper wall, there was no shock acceleration and then the structure has a double-Mach-like configuration before and after the triple point collision where a new primary strong triple point reflected off the wall.

iv. In both first and second half-cell shortly after collision, there is a local decoupling of the reaction zone from the shock front. This due to the detachment of the jet flow from the front. However, after a delay a new jet forms which reconciles the shock front to the reaction zone.

v. In both first and second half of the cell, the extended section of the transverse wave does not collide with the wall, as it converted into a system of shocklets due to its interaction with the shear layers.

vi. After collision with the upper wall, end of the first half-cell, a tongue-like unreacted pocket isolates from the front with a notch at its vertex. Hence, in the second half-cell where the shock strengthens and velocity decays the energy released by the consumption of this pocket, supports the detonation and prevent failure.

vii. In comparison to Kelvin-Helmholtz instability, the Richtmyer-Meshkov instability found to be more significant in the appearance of turbulent structures behind the front. These structures have a significant role in the consumption of the tongue-like unreacted pocket.

viii. The transverse wave does not have any significant role directly in detonation propagation. This is because the interaction of this wave with the unreacted pocket leads to the randomisation of the transverse wave. However, the simultaneous interaction of the triple point and the transverse wave with the wall, at the end of the second half-cell, gives rise to the generation of strong pressure waves. These guaranty the re-initiation of the detonation at the decaying portion of the cell cycle. Further, such interaction produces large vortex via RMI, involving the baroclinic vorticity generation, enhancing the turbulent mixing at unburnt pocket boundary.